\documentclass{article}

\usepackage[english]{babel}
\usepackage{authblk}

\usepackage[letterpaper,top=2cm,bottom=2cm,left=3cm,right=3cm,marginparwidth=1.75cm]{geometry}

\usepackage{amsmath}
\usepackage{graphicx}
\usepackage[colorlinks=true, allcolors=blue]{hyperref}
\usepackage{gensymb}

\title{A laboratory-based X-ray phase contrast microscopy system for targeting in unstained soft-tissue samples}
\author[1]{Michela Esposito}
\author[2]{Nicole Schieber}
\author[1]{Alessandro Olivo}
\author[2]{Yannick Schwab}
\author[1,*]{Marco Endrizzi}
\affil[*]{corresponding author: m.endrizzi@ucl.ac.uk}
\affil[1]{Department of Medical Physics and Biomedical Engineering, University College London, Malet Place, Gower Street, London WC1E 6BT, United Kingdom}
\affil[2]{Cell Biology and Biophysics Unit, European Molecular Biology Laboratory (EMBL), Heidelberg, 69117, Germany}
\date{}                     %% if you don't need date to appear

\begin{document}
\maketitle

\begin{abstract}
We propose an imaging system and methodology for mapping soft-tissue samples in three dimensions, with micron-scale and isotropic spatial resolution, with low-concentrations as well as in the absence of heavy metal staining. We used hard X-ray phase-contrast imaging for the X-ray ability to non-destructively probe the internal structure of opaque specimens and for enhanced contrast obtained by exploiting phase effects, even in cases with reduced or absent staining agents. To demonstrate its applicability to soft-tissue specimens, we built a compact system that is easily deployable in a laboratory setting. The imaging system is based on a conventional rotating anode X-ray tube and a state-of-the-art custom-made radiation detector. The system’s performance is quantitatively assessed on a calibration standard. Its potential for soft-tissue microscopy is demonstrated on two biological specimens and benchmarked against gold-standard synchrotron data. We believe that the approach proposed here can be valuable as a bridging imaging modality for intravital correlative light electron microscopy and be applied across disciplines where the three-dimensional morphology of pristine-condition soft tissues is a key element of the investigation. 
\end{abstract}

\section*{Introduction}
An important trend in life sciences is to integrate a variety of imaging modalities in order to combine, on the same sample, the functional information brought by fluorescence microscopy, with high-resolution sub-cellular measurements done by electron microscopy. Moreover, modern research now incorporates molecular biology assays in multicellular living model organisms, adding complexity for the integration of multi-modal imaging technologies, but enhancing the impact of the biological findings, relative to more reductionist in vitro models \cite{follain_seeing_2017}. Correlative light and electron microscopy is one way to combine functional and structural imaging on the same samples \cite{karreman_intravital_2016,durdu_luminal_2014}. For multicellular models, X-ray micro Computed Tomography (CT) has a crucial role to play in such combined workflows as it bridges the resolution gap and enables precise targeting of key features within the tissues \cite{karreman_chapter_2017,karreman_fast_2016}.
 
In comparison to X-ray attenuation contrast, which relies heavily on sample preparation \cite{meechan_crosshair_2022,stroh_situ_2021}, X-ray phase-contrast imaging offers the possibility to visualise and differentiate soft tissues even under close-to-native preparation conditions.  Recent studies \cite{feinauer_local_2021,zhang_sample_2022,bosch_functional_2022} have shown how microCT can play a key role in combining intravital correlative light and electron microscopy (CLEM). MicroCT uniquely revealed anatomical features of the tissues as seen in the resin embedded material, ready to be observed by EM. Providing enhanced precision, in the order of a few microns, the targeting by EM of the region of interest was dramatically accelerated, thus enabling multiple observations. Other fixation methods, adapted to small living specimens and tissues, consist in arresting the samples by high pressure freezing (HPF) \cite{mcdonald_review_2009}. The vitrified samples are processed and embedded in a resin by freeze substitution (FS) to be then imaged by EM at room temperature. Such approaches are very powerful to preserve good ultrastructure together with the immunogenicity of the tissue, i.e. the possibility to apply affinity probes to localize and quantify protein expression at specific subcellular position \cite{studer_electron_2008,korogod_ultrastructural_nodate,tsang_high-quality_2018}. Such samples can have very low contrast, dependent on the freeze substitution mixture, thus precluding the possibility to target regions of interest with conventional microCT. It is therefore desirable to be able to work at reduced concentrations of staining, as well as avoiding it completely, however preserving sufficient contrast in the images. This would enable an even more flexible combination of imaging techniques, for example avoiding quenching fluorescence, and provide a pathway for visualising and differentiating multi-cellular organisms' soft tissues in close-to-native preparation. This could be exploited to provide a solution for a current gap in multi-scale biological imaging: the precise targeting of key features within the tissue. Bridging this gap would be beneficial both in terms of simplifying and speeding up data collection and analysis, and in terms of opening new ways to investigate the samples. This procedure requires the inspection of relatively large samples, at the cubic millimetre scale, with a resolution of the order of one micron.  Developing efficient CLEM workflows and targeting methods to link functional imaging to ultrastructure is of prime importance - specially considering the rise of intravital imaging to tackle the fundamental mechanisms of life especially in integrated and multicellular models. 

We propose a laboratory-based system with a relatively compact footprint and that operates by combining a rotating anode X-ray tube and a state-of-the-art image receptor. The system has been designed around the need of obtaining three-dimensional representations of approximately one cubic millimetre of soft tissue samples, with micron-scale spatial resolution. We report on the instrument characterisation, the optimal trade-offs between sensitivity and spatial resolution and compare its capabilities to achieve targeting-without-staining to gold-standard synchrotron radiation micro-tomography datasets. 
\begin{figure}[htbp]
\centering \includegraphics[width=1\columnwidth]{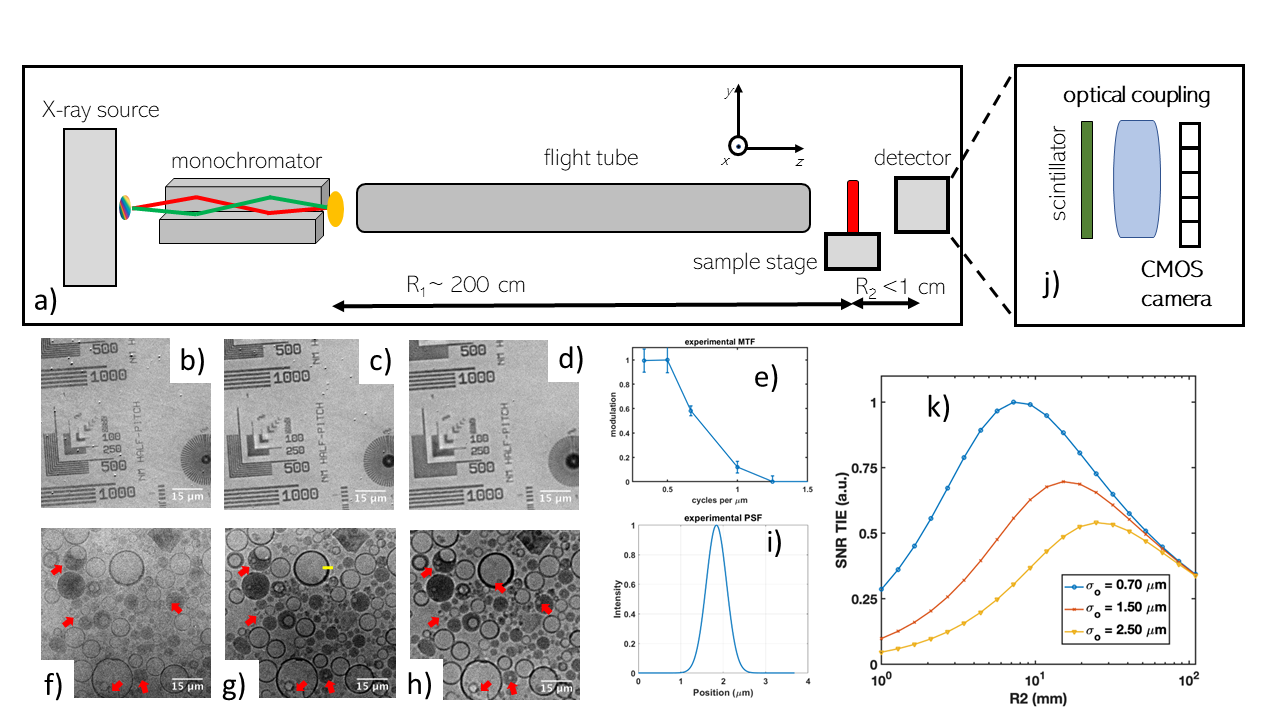}
\caption{\label{fig:setup} \textit{a}) Schematic of the x-ray imaging system including a rotating anode x-ray source, a doubly-bent multi-layer monochromator, a vacuum flight tube, a sample stage, and an imaging detector. Details of the component of the custom-made detector are shown in panel \textit{j}). Optimisation of the propagation distance: \textit{b}) - \textit{d}) A resolution target imaged at a propagation distance R$_2$ of approximately 0, 2, 4 mm, respectively. \textit{e}) Modulation Transfer Function (MTF) calculated from the resolution target at $R_2$=2 mm. \textit{f}) – \textit{h}) A contrast test object comprising glass microspheres imaged at a propagation distance  $R_2$  of approximately 0, 2 and 4 mm, respectively. The red arrowheads highlight details in the images that are hardly, if at all, visible in attenuation contrast and become increasingly clearer at longer propagation distances. \textit{i}) Point Spread Function (PSF) calculated from an edge profile (indicated by a yellow line) in the image at  $R_2$=2 mm. \textit{k}) Theoretical prediction of the SNR as a function of the propagation distance $R_2$from Equation \ref{eq:SNR}.}
\end{figure}
\section*{Methods}
X-ray phase-contrast imaging \cite{paganin_coherent_2006} offers a unique approach for visualising soft tissue with high resolution in three-dimensions, even when the conventional attenuation contrast does not provide a good enough signal to differentiate structures. One of the simplest set-ups for exploiting phase-effects in X-ray imaging is the so-called in-line or free-space propagation Figure \ref{fig:setup}\textit{a}, where a distance is introduced between the sample and the detector to allow the wavefield to evolve before detection \cite{wilkins_phase-contrast_1996}. Perturbations to the phase of the wavefield imparted by the object are in this way translated into intensity modulations that can be detected and interpreted. For sufficiently slowly varying fields, this can be described as phase contrast by defocus
\begin{equation}
I(x,y,R_2)=I(x,y,R_1)\big[ 1-\nabla^2_\perp\Phi(x,y,R_1)\big]
\label{eq:TIE}    
\end{equation}
where a paraxial wavefield $\Psi(x,y,z)$ is written in terms of its intensity $I\equiv |\Psi(x,y,z)|^2 $ and phase $\Phi(x,y,z)=arg(\Psi(x,y,z))$, $\lambda$ is the radiation wavelength and $\nabla^2_\perp$ is the Laplacian in the (x,y) plane. Equation \ref{eq:TIE} describes how the beam intensity evolves propagating from $R_1$ to $R_2$  and shows how phase contrast by defocus increases linearly with the propagation distance and depends on the Laplacian of the phase. The validity of Equation \ref{eq:TIE} is subject to propagation distances being sufficiently small and for sufficiently slowly varying intensities in $x$ and $y$. For a laboratory-based free-space propagation systems a with finite source sizes and longer propagation distances, there is also the competing effect where the blurring introduced by the source distribution increasingly washes away the intensity modulations due to phase effects. The signal-to-noise ratio (SNR) can be expressed analytically for an edge-like simple feature \cite{gureyev_simple_2008}
\begin{equation}
SNR \propto \frac{R_t(M-1)}{M^{1/2}[M^2 \sigma_0^2+(M-1)^2 \sigma_s^2+\sigma_d^2]}
\label{eq:SNR}
\end{equation}
where $R_t$  is the distance between the X-ray source and the detector, $M=R_t/R1$ is the geometrical magnification given by the ratio between $R_t$   and the source to sample distance $R_1$, and $\sigma_{(0,d,s)}$ are characteristic lengths of the edge-like object, and the source and detector point spread functions, respectively. The plot in Figure \ref{fig:setup}\textit{k} shows the SNR in arbitrary units as a function of the distance $R_2$ between the sample and the detector for different values of the detector point spread function ($\sigma_d$). This plot shows the optimal propagation distances for our system, as well as the motivation for pushing the spatial resolution of the detector in the sub-micron region.

\subsection*{X-ray imaging system}
A schematic overview of the x-ray imaging system is shown in Figure \ref{fig:setup}\textit{a}). X-rays are produced by a Rigaku Multi-Max 9 source, using a rotating Cu anode operated at 46 Kvp and 26 mA.  Doubly-bent multi-layer optics (Varimax Cu-HF) select the Cu $K_\alpha$ lines (8.05 and 8.04 keV) with 1\% resolution and focus the beam to a 350 $\mu$m focal spot. We note that a monochromatic beam, although not strictly necessary for propagation-based phase contrast imaging \cite{wilkins_phase-contrast_1996}, allows for quantitative analysis \cite{brombal_monochromatic_2019}. The 2.2-m long flight path ($R_1$) from the optics focal point to the sample, is realised in vacuum to maximise the X-ray flux at the sample position. A 5-degree-of-freedom sample stage was used for sample positioning and rotation. The detector was custom-built, comprising a garnet scintillator, interchangeable objectives, a tube lens, and a CMOS pixel sensor featuring 1608×1608 11-$\mu$m pixel pitch. Objectives providing a 10× and 20× magnification were chosen, giving an effective pixel size of 1.1 and 0.55 $\mu$m and field-of-view (FOV) of 1.8×1.8 and 0.9×0.9 mm$^2$, respectively. A schematic overview of the detector components is reported in Figure \ref{fig:setup}\textit{j}).  

The choice of the propagation distance ($R_2$) results from the trade-off between phase sensitivity and spatial resolution. For this purpose, we used a calibration standard made of Au bars deposited on a SiN$_3$ membrane. The modulation at each spatial frequency was measured and used to characterise the Modulation Transfer Function (MTF) of the system (\ref{fig:setup}\textit{b-e}). Additionally, a test phantom featuring glass microspheres on a kapton membrane, was used to visualise how changing the propagation distance affects the contrast. The latter were obtained through the differentiation of edge profiles, giving the system Point Spread Function (PSF).
\begin{figure}[htbp]
\centering \includegraphics[width=0.5\columnwidth]{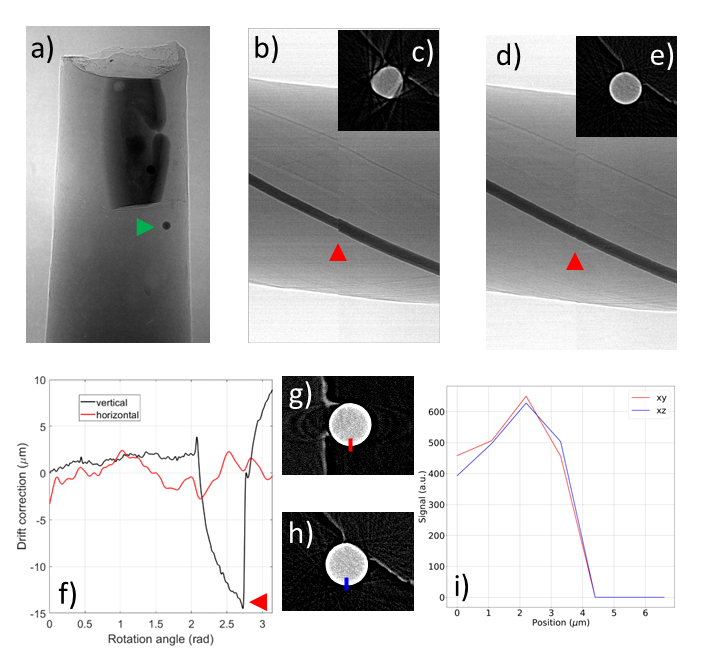}
\caption{\label{fig:tracking} Data pre-processing for drifts correction. \textit{a}) Projection image of the low-staining protocol drosophila sample. The fiducial marker used to correct system drifts and vibrations is highlighted by a green arrowhead. \textit{b}) Section of the sinogram with a red arrowhead highlighting the issue with system drifts and instabilities, and \textit{c}) the resulting reconstruction artefacts. \textit{d}) Same section of sinogram after drift correction and \textit{e}) effect on the correction on the reconstruction. \textit{f}) Measured drift values for the vertical and horizontal coordinates. A large drift is observed vertically (red arrowhead). The corresponding point in the raw and corrected sinogram is marked by red arrowheads in panels \textit{b}) and \textit{c}). \textit{g}) - \textit{h}) Two reconstructed orthogonal slices containing the fiducial marker, showing effectiveness of the correction and isotropic spatial resolution, also quantitatively compared with two intensity profiles in panel \textit{i}). }
\end{figure}
\subsection*{Biological samples}
The biological samples chosen for this work were two drosophila larvae, a widely used model in research. The sample were fixed by high pressure freezing (HPF: 2000 bars at -196$\degree$C) and processed by freeze substitution (FS at -90$\degree$C in acetone + 0.1\% uranyl acetate, rinsed in acetone and infiltrated in lowicryl HM20 at -45$\degree$C, polymerization by UV light at -45$\degree$C). Note that at these concentrations, uranyl acetate minimally interferes with the immunogenicity of the tissues or with the fluorescence of fluorescent proteins \cite{nixon_single_2009,kukulski_correlated_2011}.
\subsection*{Acquisition details}
Two sample of drosophila were chosen for laboratory microCT imaging. One was prepared without heavy metal staining and imaged with an effective pixel size of 0.55 $\mu$m and 0.9×0.9 mm$^2$ field of view. Due to its transverse dimension being larger than the FOV, the sample was imaged over 360$\degree$ with the rotation axis placed at the edge of the FOV. Opposite projection images were then registered together and merged to provide a full image of the sample over 180$\degree$. 5000 projection images were acquired with an exposure time of 40 s and an angular increment of 0.072$\degree$. The other sample was prepared with low concentration of metal staining (0.1\%UA in acetone) was imaged with an effective pixel size of 1.1 $\mu$m and 1.8×1.8 mm$^2$ field of view. 2500 projections with 30 s exposure time were acquired over 180$\degree$. Both samples were additionally imaged at Diamond Light Source on the I13-2 beamline, using 12 keV X-ray energy and an effective pixel size of 0.8 $\mu$m, for obtaining the comparison gold-standard. Intensity projection were retrieved using the so-called Paganin’s phase retrieval algorithm \cite{paganin_simultaneous_2002} and reconstructed using the dedicated data processing pipeline at the beamline \cite{atwood_high-throughput_2015}.

\subsection*{Reconstruction}
The first step in the reconstruction of laboratory-based microCT datasets acquired with our lab-based system was to address system instabilities that were manifest in the sinograms and resulted in reconstruction artefacts (Figure 2a and b). This was achieved by using fiducial markers to track the position of the sample in the detector plane and correct for drifts. Glass (SiO$_2$) microspheres with a 25-$\mu$m radius were attached to the samples, as shown in Figure \ref{fig:tracking}\textit{c} (green arrow). The position of the spheres, at each projection angle, was calculated using a built-in Matlab function based on the Hough transform. Sample drift was then estimated by comparing the measured sphere’s trajectory with the expected one, i.e., constant in the vertical direction and sinusoidal in the horizontal one. Figure \ref{fig:tracking}\textit{a} and \ref{fig:tracking}\textit{d} show a section of the same sinogram, before and after correction, respectively. Artefacts arising from drifts can be seen comparing panels b and e where reconstructed slices containing the fiducial markers are shown before and after drift corrections, respectively. The estimated drift is shown in Figure \ref{fig:tracking}\textit{f} for the low-stained drosophila sample. A large vertical drift is visible and has been highlighted in both raw and corrected sinograms with a red arrowhead. Two orthogonal slices containing the fiducial markers are reported in panels g and h, with corresponding intensity profiles across the sphere edges shown in panel i. After processing the system correctly represents the marker with isotropic spatial resolution. Following drift correction, intensity projections were phase-retrieved using the Paganin’s phase retrieval \cite{paganin_simultaneous_2002}, assuming that $\beta$ and $\delta$ are proportional through the sample. This corresponds to the hypothesis of the sample being constituted of a single material, although not strictly true in most cases, this has proven a practical and valid approach, especially for soft tissues. The sinograms extracted from phase-retrieved projection were then reconstructed using a standard Filtered Back Projection algorithm.

\begin{figure}[htbp]
\centering \includegraphics[width=1\columnwidth]{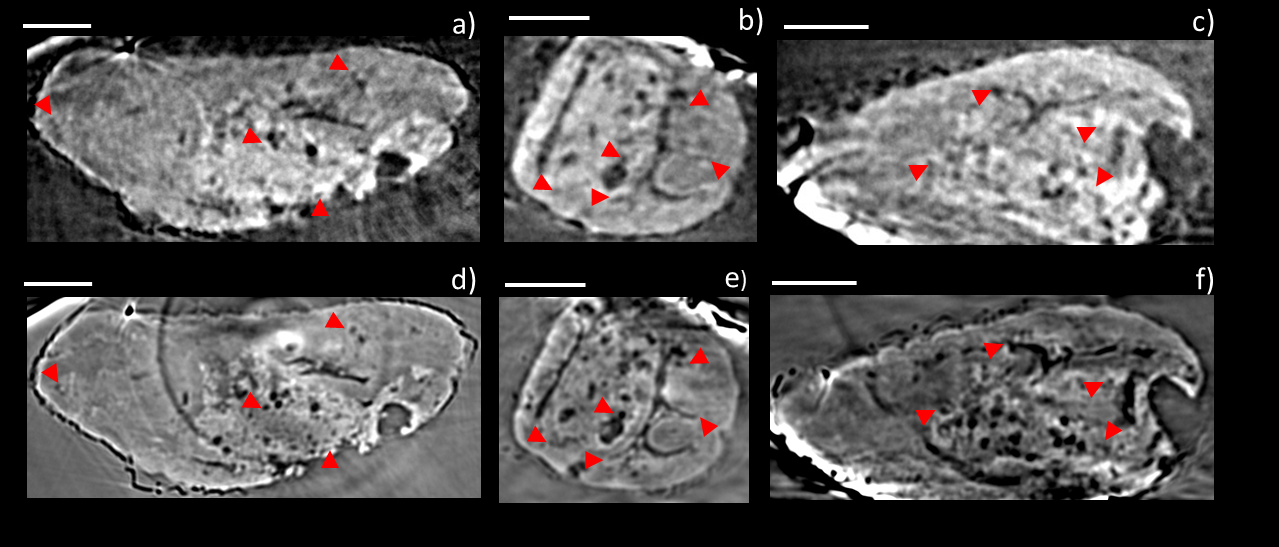}
\caption{\label{fig:unstained} Tomographic slices of the unstained drosophila larva from laboratory (\textit{a}-\textit{c}) and synchrotron (\textit{d}-\textit{f}) scans. Red arrowheads highlight specific biological features in both datasets.  Scale bar 50 $\mu$m.}
\end{figure}
\begin{figure}[htbp]
\centering \includegraphics[width=1\columnwidth]{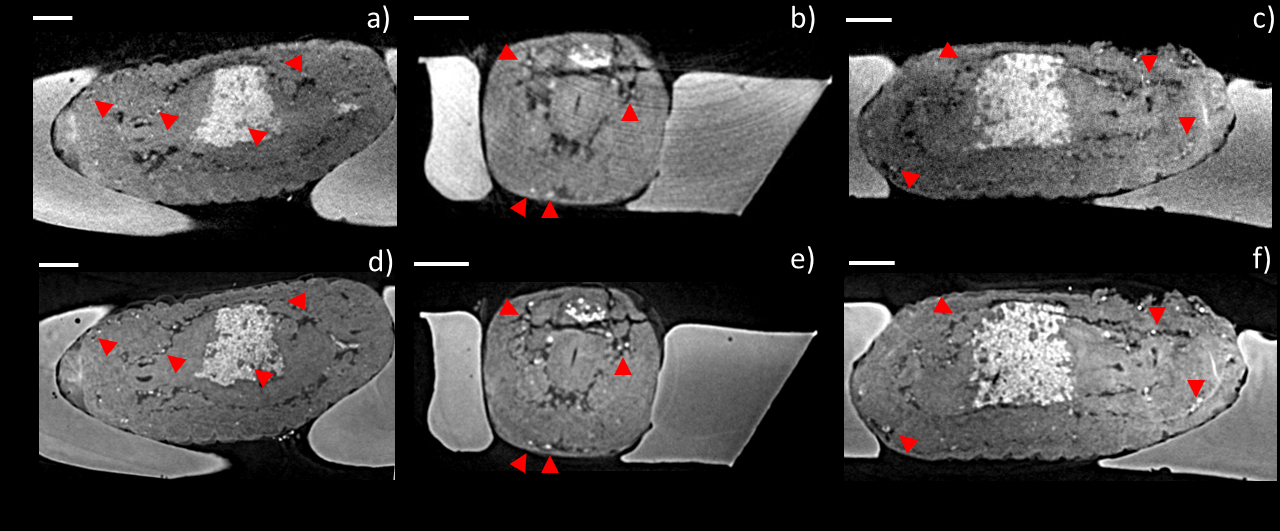}
\caption{\label{fig:stained} Tomographic slices of the low-staining drosophila larva from laboratory (\textit{a}-\textit{c}) and synchrotron (\textit{d}-\textit{f}) scans. Red arrowheads highlight specific biological features in both datasets.  Scale bar 50 $\mu$m.}
\end{figure}
\section*{Results and Discussion}
The resolution target and the contrast phantom enabled to find the working point of our imaging system by simultaneously optimising spatial resolution and phase contrast. Figure\ref{fig:setup}\textit{b-d}) and f-h) show details of the resolution target and test object respectively, for propagation distances of approximately 0, 2 and 4 mm, respectively. Phase contrast increases with propagation distance, as it is mostly visible for weakly absorbing features (highlighted by red arrows in panels f-h)), in line with expectations from Equation \ref{eq:TIE}. However, increasing the propagation distance also results in a loss of spatial resolution., as shown in panels b-d). A trade-off between spatial resolution and phase contrast is therefore necessary, leading to the choice of a propagation distance of 2 mm, corresponding to the images in panels c) and g). At the choice propagation distance, the system resolution was evaluated via the MTF (panel e)) and LSF (panel i)), suggesting a system resolution of around 1 $\mu$m. The plots in Figure \ref{fig:setup}\textit{k} show how the maximum SNR in the phase-contrast image shifts when the spatial resolution of the system becomes lower and offers a look up table to quickly adjust the system when pixel size has to be changed, for example for accommodating a larger field of view.
Figure \ref{fig:unstained} presents three views of the drosophila sample prepared without staining: coronal, axial, and sagittal tomographic slices obtained with our laboratory system are shown in panels \textit{a-c}). Gold-standard reconstructions from synchrotron data are shown for comparison in panels \textit{d-f}) The two datasets (laboratory and synchrotron) were manually aligned, and the views presented are for the same locations. Red arrowheads highlight anatomical landmarks that can be recognised in both datasets. Although our compact system could not fully match the image quality achievable with synchrotron radiation, it still provided a volumetric representation with enough detail to enable targeting of smaller areas within the volume that was inspected.

\begin{figure}[htbp]
\centering \includegraphics[width=0.5\columnwidth]{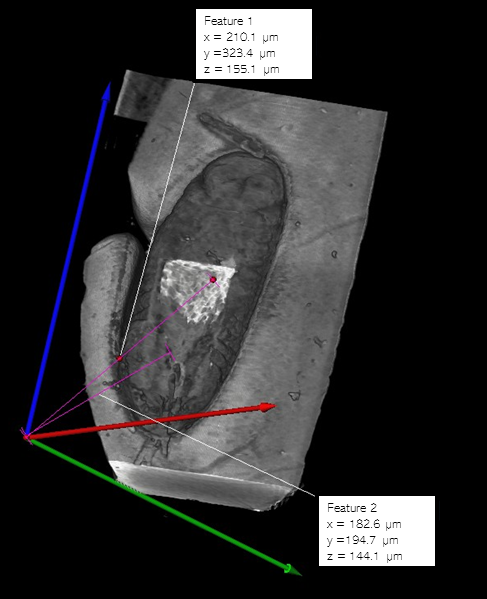}
\caption{\label{fig:3D} Exemplar targeting capabilities of the laboratory system. Two features of interest are highlighted in the reconstructed volume and their position, with the three cartesian axes is displayed. Coordinates for the features of interest are measured with an uncertainty of $\pm$1.1 $\mu$m.}
\end{figure}

A similar comparison is provided in Figure \ref{fig:stained} for the drosophila sample prepared with the low-concentration heavy metal staining.  Panels \textit{a-c}) are obtained from laboratory scans and panels \textit{d-f}) from synchrotron ones. Also here, we have highlighted a few anatomical landmarks that are visible in both reconstructions.
We have found that spatial resolution and sensitivity of our custom laboratory system allows the targeting, with micrometric precision, of features of interest within the micro-organism. Results from an exemplar targeting workflow are provided in Figure \ref{fig:3D}. Two features of interest were chosen and marked within the sample. Their position is subsequently measured with respect to a three-dimensional coordinate system with origin in one corner of the resin block, that served as a benchmark point between the virtual slicing offered by the microCT dataset against the actual sample. The uncertainty in the position of the feature of interest is associated with the voxel dimensions, which in this case are of 1.1 µm isotopically.

\section*{Conclusion}
We have presented a compact, laboratory scale X-ray microscope for three-dimensional imaging of soft tissue, that was tuned to inspect approximately one cubic millimetre with micrometric spatial resolution. The system was characterised and optimised for phase-contrast X-ray microCT on a calibration standard and a contrast phantom. It was subsequently tested on two samples of drosophila larva, a widely used multicellular organism in developmental biology. The quality of the results was assessed against gold-standard synchrotron data, with comparable results in terms of the ability to localised anatomical landmarks within the volumes of interest. We have found the system capable of delivering high-quality images in both a reduced-staining and unstained soft tissue samples, within which it was possible to measure, with micron accuracy, the three-dimensional position of anatomical landmarks. We believe that having achieved 3D micron-scale imaging of unstained tissue, within a compact and easily deployable instrument, is an important enabling paradigm towards the long-term vision of a seamless structure-function analysis workflow on samples from live-cell imaging to cryo-electron microscopy.
\section*{Acknowledgements}
We gratefully acknowledge Diamond Light Source for time on Beamline I-13 under Proposal MT17971-2. This project has received funding from the ATTRACT project funded by the EC under Grant Agreement 777222. This work was supported by the Wellcome Trust 221367/Z/20/Z and by the National Research Facility for Lab X-ray CT (NXCT) through EPSRC grant EP/T02593X/1. This work was supported by the EPSRC through grants EP/M028100/1 and EP/P023231/1. Research reported in this publication was supported by the National Institute of Biomedical Imaging and Bioengineering of the National Institutes of Health under Award Number R01EB028829. The content is solely the responsibility of the authors and does not necessarily represent the official views of the National Institutes of Health.
\bibliographystyle{unsrt}
%\bibliography{attract}

\end{document}